\documentclass[sigconf, authorversion,nonacm]{acmart}

\copyrightyear{2025}
\acmYear{2025}
\setcopyright{acmlicensed}\acmConference[SURE '25]{Proceedings of the 2025 Workshop on Software Understanding and Reverse Engineering}{October 13--17, 2025}{Taipei, Taiwan}
\acmBooktitle{Proceedings of the 2025 Workshop on Software Understanding and Reverse Engineering (SURE '25), October 13--17, 2025, Taipei, Taiwan}
\acmDOI{10.1145/3733822.3764670}
\acmISBN{979-8-4007-1910-3/2025/10}

\usepackage[all]{nowidow}
\usepackage{siunitx}
\usepackage{multirow}
\usepackage{amsmath}
\usepackage{xspace}
\usepackage{graphicx}
\usepackage{enumitem}
\usepackage[linesnumbered,ruled,vlined]{algorithm2e}
\usepackage{caption}
\usepackage{subcaption}
\usepackage{colortbl}
\usepackage{diagbox, eqparbox, hhline}

\usepackage{booktabs}

\newcommand{\aref}[1]{\hyperref[#1]{Appendix~\ref*{#1}}}

\newcommand{\eg}{e.g.,}

\newcommand{\shortsection}[2][.]{\vspace{1mm}\noindent\textbf{#2#1}}

\newif{\ifanonymous}
\anonymousfalse{} 

\newcommand{\libiht}{\textsc{LibIHT}\xspace}

\usepackage{pifont}
\newcommand{\cmark}{\ding{51}}%
\newcommand{\xmark}{\ding{55}}%

\usepackage{framed}
\usepackage{xcolor}

\definecolor{shadecolor}{RGB}{240,240,240}

\newenvironment{takeaway}[1]{
  \begin{shaded}
  \noindent\textbf{Takeaway: #1}\par
}{
  \end{shaded}
}

\begin{document}

\title[]{\libiht: A Hardware-Based Approach to Efficient and Evasion-Resistant Dynamic Binary Analysis}


\author{Changyu Zhao}
\orcid{0009-0009-9375-1519}
\affiliation{%
  \institution{University of Wisconsin-Madison}
  \city{Madison}
  \state{Wisconsin}
  \country{USA}}
\email{changyuz@stanford.edu}

\author{Yohan Beugin}
\orcid{0000-0003-0991-7926}
\affiliation{%
  \institution{University of Wisconsin-Madison}
  \city{Madison}
  \state{Wisconsin}
  \country{USA}}
\email{ybeugin@cs.wisc.edu}

\author{Jean-Charles Noirot Ferrand}
\orcid{0009-0009-9650-4011}
\affiliation{%
  \institution{University of Wisconsin-Madison}
  \city{Madison}
  \state{Wisconsin}
  \country{USA}}
\email{jcnf@cs.wisc.edu}

\author{Quinn Burke}
\orcid{0000-0003-1719-3112}
\affiliation{%
  \institution{University of Wisconsin-Madison}
  \city{Madison}
  \state{Wisconsin}
  \country{USA}}
\email{qkb@cs.wisc.edu}

\author{Guancheng Li}
\orcid{0009-0007-4226-656X}
\affiliation{%
  \institution{Tencent Xuanwu Lab}
  \city{Beijing}
  \state{Beijing}
  \country{China}}
\email{atumli@tencent.com}

\author{Patrick McDaniel}
\orcid{0000-0003-2091-7484}
\affiliation{%
  \institution{University of Wisconsin-Madison}
  \city{Madison}
  \state{Wisconsin}
  \country{USA}}
\email{mcdaniel@cs.wisc.edu}

\begin{abstract}
Dynamic program analysis is invaluable for malware detection, debugging, and performance profiling. 
However, software-based instrumentation incurs high overhead and can be evaded by anti-analysis techniques. 
In this paper, we propose \libiht, a hardware-assisted tracing framework that leverages on-CPU branch tracing features (Intel Last Branch Record and Branch Trace Store) to efficiently capture program control-flow with minimal performance impact. 
Our approach reconstructs control-flow graphs (CFGs) by collecting hardware generated branch execution data in the kernel, preserving program behavior against evasive malware. We implement \libiht as an OS kernel module and user-space library, and evaluate it on both benign benchmark programs and adversarial anti-instrumentation samples. 
Our results indicate that \libiht reduces runtime overhead by over 150× compared to Intel Pin (7× vs 1,053× slowdowns), while achieving high fidelity in CFG reconstruction (capturing over 99\% of execution basic blocks and edges).
Although this hardware-assisted approach sacrifices the richer semantic detail available from full software instrumentation by capturing only branch addresses, this trade-off is acceptable for many applications where performance and low detectability are paramount. Our findings show that hardware-based tracing captures control flow information significantly faster, reduces detection risk and performs dynamic analysis with minimal interference.
\end{abstract}


\keywords{Hardware; Tracing; Malware; Reverse engineering; Binary analysis}

\maketitle{}

\section{Introduction}\label{introduction}

Dynamic binary program analysis plays a critical role in reverse engineering and security evaluation~\cite{nielson_principles_2015, alrabaee_binary_2020, ball_concept_1999, nethercote_dynamic_2004}, enabling analysts to understand program behavior when source code is unavailable. In security contexts, such analysis is indispensable for identifying vulnerabilities, detecting malware, and verifying that system components adhere to expected behavior. Reverse engineering not only facilitates the discovery of hidden or obfuscated code constructs but also informs the development of robust defensive strategies.

Despite the importance of dynamic analysis, existing tools typically rely on software-based instrumentation methods (\eg Intel Pin~\cite{luk_pin_2005} and DynamoRIO~\cite{bruening_efficient_2004}) that inject monitoring code into the target program~\cite{nethercote_valgrind_2007, feiner_comprehensive_2012, scott_retargetable_2003, bhansali_framework_2006}. While these approaches provide fine-grained visibility into runtime behavior, they incur substantial performance overhead and often perturb the program’s natural execution. Furthermore, because software-based instrumentation inserts hooks and modifies the target binary at runtime, these frameworks are routinely subject to anti‐analysis and evasion techniques~\cite{polychronakis_measuring_2017, filho_evasion_2022, delia_evaluating_2022, delia_sok_2019, zhechev_security_2018, park_detection_2024, filho_reducing_2020}, which in turn compromise control-flow reconstruction and the overall analysis reliability.

In this paper, we introduce \libiht, a hardware-assisted tracing framework that complements traditional dynamic binary instrumentation by addressing its performance and detectability limitations. Rather than injecting instrumentation code at runtime, \libiht leverages on-CPU branch tracing features built into commodity Intel x86/x64 processors~\cite{noauthor_intel_nodate}. Specifically, \libiht uses Last Branch Record (LBR) and Branch Trace Store (BTS) to capture program execution with minimal overhead. By operating directly at the hardware level, \libiht significantly reduces performance penalties and preserves natural execution. 
Although this approach records only branch addresses (thus offering lower granularity than full software instrumentation), we hypothesize that such coarse-grained data is sufficiently accurate for effective reverse engineering and dynamic analysis. Notably, while existing hardware-assisted tools for enforcing control flow integrity (CFI) achieve high accuracy and resilience in security enforcement, their designs have predominantly targeted runtime protection rather than assisting reverse engineering~\cite{willems_down_2012, cheng_ropecker_2014, pappas_transparent_2013, zhou_hardstack_2019, liu_retrofitting_2022}. The complexity of harnessing hardware-level tracing and associated trade-offs have thus far prevented a unified solution tailored for reverse engineering. 
As a result, \libiht is an ideal solution when analysis scenarios demand high performance and low detectability more than exhaustive semantic information.

We evaluate \libiht across a diverse set of binaries, including both standard and adversarial executables, assessing its effectiveness in (1) reconstructing accurate control flow graphs, (2) minimizing runtime overhead compared to software-based instrumentation, and (3) maintaining resilience against common anti-analysis techniques. By leveraging the processor’s built-in tracing features, \libiht records execution data at the basic-block level and reconstructs precise control flow with minimal interference. Our results show that \libiht reconstructs control flow with over 98\% accuracy—achieving a mean Jaccard Index of 0.9836, 99.48\% block coverage, and 99.69\% edge coverage—while reducing average runtime overhead to 7× (versus 237× for DynamoRIO and 1,053× for Intel Pin). In other words, \libiht delivers high-fidelity analysis at a fraction of the cost: it eliminates intrusive instrumentation and thereby avoids the severe slowdowns of traditional tools. Furthermore, its reliance on hardware-level tracing enables robust analysis that bypasses the diverse set of anti-debugging and anti-instrumentation benchmarks we considered.

Our contributions are as follows:
\begin{itemize}
    \item We quantify challenges faced by software-based dynamic analysis by measuring both performance overhead and susceptibility to evasion across real-world benchmarks.
    \item We introduce \libiht, a hardware-assisted program analysis framework that leverages Intel LBR and BTS to enable efficient and accurate execution tracing.
    \item We evaluate \libiht on a diverse set of binaries, demonstrating its effectiveness in reconstructing control flow graphs while maintaining low runtime overhead and resilience fight against anti-analysis techniques.
\end{itemize}

The complete \libiht toolchain (kernel module and user-space library) is released as open-source software at the following URL:\url{https://github.com/libiht/libiht}

\section{Background}\label{background}

\subsection{Software-Based Dynamic Analysis Tools}\label{background:software-analysis}
\begin{figure}[ht!]
    \centering
    \includegraphics[width=.95\columnwidth]{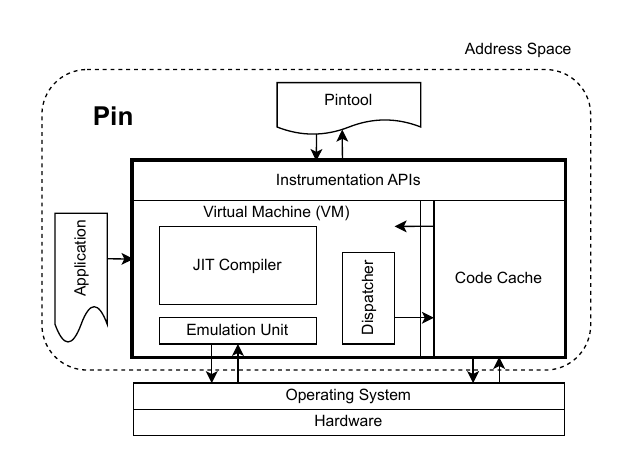}
    \caption{High-level architecture of the Intel Pin (adapted from~\cite{luk_pin_2005}). The tool inserts a virtual machine, JIT compiler, and dispatcher to dynamically transform the target program, storing instrumented code in a dedicated code cache.}
    \Description{Diagram of Intel Pin architecture showing an application interacting with a virtual machine that contains a JIT compiler, emulation unit, dispatcher, and code cache. Instrumentation APIs link to user-defined Pintools, and the instrumented code is executed from the cache.}
    \label{fig:pin}
\end{figure}

\begin{figure}[ht!]
    \centering
    \includegraphics[width=.95\columnwidth]{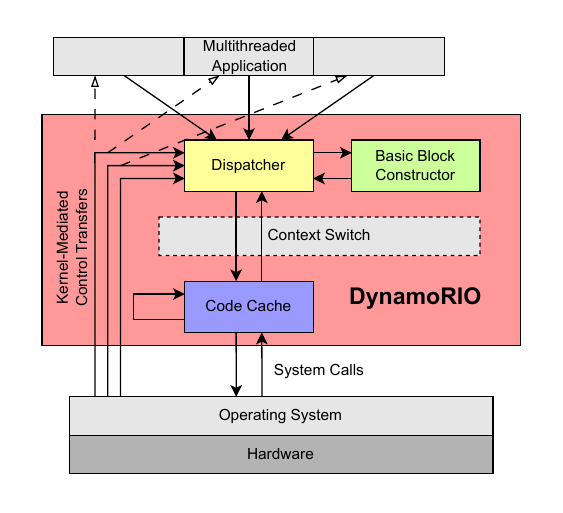}
    \caption{Overview of DynamoRIO’s architecture (adapted from~\cite{bruening_efficient_2004}). Similar to Intel Pin, it uses a code cache for instrumented basic blocks, with a dispatcher managing control transfers to and from the running application.}
    \Description{Diagram of DynamoRIO architecture showing a multithreaded application interacting with DynamoRIO. Inside DynamoRIO, a dispatcher manages execution, a basic block constructor creates instrumented code fragments, and a code cache stores these blocks. Control transfers, system calls, and context switches connect the application, operating system, and hardware.}
    \label{fig:dynamorio}
\end{figure}

Traditional dynamic analysis is typically performed with dynamic binary instrumentation (DBI) frameworks such as Intel Pin~\cite{luk_pin_2005} and DynamoRIO~\cite{bruening_efficient_2004}. These DBI tools work by injecting monitoring code into the target program at runtime to capture execution traces and reconstruct control flow. \autoref{fig:pin} and \autoref{fig:dynamorio} illustrate the high-level designs of Pin and DynamoRIO, respectively. Both frameworks employ just-in-time (JIT) compilation and maintain a dedicated code cache to hold instrumented code blocks, thereby enabling detailed runtime monitoring. 
Although both Intel Pin and DynamoRIO support flexible instrumentation and fine-grained tracing, they suffer from two major drawbacks:
\begin{itemize}
  \item \textbf{Performance Overhead}: Instrumentation can slow down execution by orders of magnitude, making it impractical for real-time or large-scale analysis.
  \item \textbf{Anti-Analysis Evasion}: Many programs, especially malware, employ anti-instrumentation techniques to detect and evade software-based monitoring~\cite{polychronakis_measuring_2017, filho_evasion_2022, delia_evaluating_2022, delia_sok_2019, zhechev_security_2018, park_detection_2024, filho_reducing_2020}.
\end{itemize}

\begin{table}[ht]
\centering
\caption{Instrumentation overhead for some commonly used \texttt{coreutils} under Intel Pin and DynamoRIO. Each workload command is listed in \aref{appendix:workload-commands}. All timings are in milliseconds; slowdown factors are shown in parentheses.}
\label{tab:instr_overhead}
\begin{tabular}{lccc}
\toprule
Program & Native & Intel Pin & DynamoRIO \\
\midrule
\texttt{ls}   & 7 ms & 829 ms (118×) & 94 ms (13×) \\
\texttt{dd}   & 14 ms & 625 ms (45×)  & 68 ms (5×)   \\
\texttt{echo} & 4 ms & 508 ms (127×) & 54 ms (14×)  \\
\texttt{sort} & 9 ms & 600 ms (67×)  & 63 ms (7×)   \\
\texttt{wc}   & 8 ms & 559 ms (70×)  & 56 ms (7×)   \\
\texttt{cat}  & 7 ms & 497 ms (71×)  & 55 ms (8x)   \\
\bottomrule
\end{tabular}
\end{table}

\shortsection{Limitations}\label{background:software-analysis-limitations}
Although these binary instrumentation frameworks are widely used and effective for detailed runtime program analysis, their design inherently prioritizes fine-grained visibility and flexible instrumentation over execution performance. As our motivating experiment results demonstrate (see \autoref{tab:instr_overhead}), even simple, deterministic workloads such as \texttt{ls}, \texttt{dd}, \texttt{echo}, \texttt{sort}, \texttt{wc}, and \texttt{cat} (see \aref{appendix:workload-commands}) suffer from significant performance degradation, even without adopting any analysis logic to the binary. For example, while native execution of \texttt{echo} requires only 4~ms, running it under Intel Pin increases the runtime to 508~ms, representing a slowdown of over 127×. Similar trends are observed with the other commands, highlighting that the overhead introduced by these frameworks can severely distort execution timing. This trade-off renders them impractical for performance-sensitive or real-time systems and may lead to inaccuracies in analyses such as control flow graph (CFG) reconstruction, particularly in security-critical applications where fidelity and stealth are paramount.

\begin{figure*}[ht!]
    \centering
    \includegraphics[width=\textwidth]{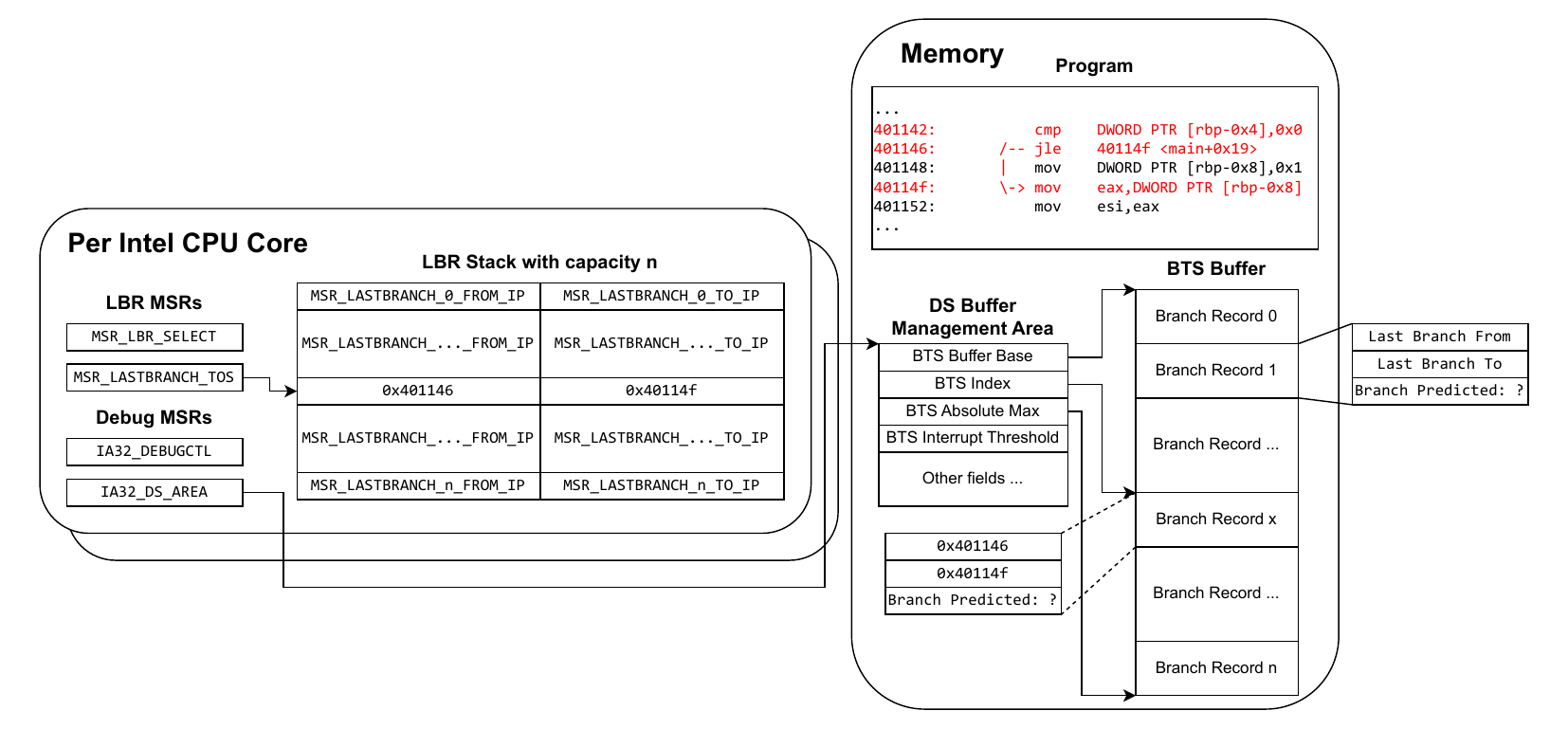}
    \caption{High-level architecture of Intel's LBR and BTS mechanism. LBR maintains a stack of MSRs on each CPU core for the most recent branches, while BTS extends this by logging branch records into a dedicated buffer in memory.}
    \Description{Diagram of Intel’s hardware tracing. On each CPU core, Last Branch Record (LBR) MSRs store the source and target addresses of recent branches in a stack. Debug MSRs configure tracing. A program in memory executes instructions, and branch information is written to the DS buffer management area, which tracks buffer base, index, and thresholds. The Branch Trace Store (BTS) buffer contains multiple branch records, each recording the last branch taken, its destination, and whether it was predicted.}
    \label{fig:hw_trace}
\end{figure*}

\subsection{Hardware-Assisted Tracing Mechanisms}\label{background:hardware-tracing}
To address the limitations of software-based tracing, modern x86 processors provide hardware-level branch tracing mechanisms. These include \textit{Last Branch Record (LBR)}, \textit{Branch Trace Store (BTS)}, and \textit{Intel Processor Trace (Intel PT)}~\cite{noauthor_intel_nodate}. By capturing control-flow transitions at the processor level, these features enable efficient tracing without instrumenting program code.

\shortsection{LBR}\label{background:lbr}
\autoref{fig:hw_trace} (left) illustrates the Last Branch Record mechanism.  LBR uses a small, fixed‐size circular buffer of Model Specific Registers (MSRs) to record the source and destination addresses of the most recent branch instructions executed by the CPU. Each new branch pushes the oldest entry out of the buffer, so LBR always reflects the last $n$ (ranging from 4 to 32 base on CPU model) branches taken.  Because all operations occur entirely in hardware registers, enabling LBR incurs virtually no runtime overhead. This makes LBR ideal for capturing short‐term execution snapshots, such as profiling tight code paths or quickly sampling control‐flow behavior without disrupting program throughput.

\shortsection{BTS}\label{background:bts}
\autoref{fig:hw_trace} (right) shows the Branch Trace Store mechanism. Instead of limited MSRs, BTS writes every branch event into a dedicated in‐memory trace ring buffer managed by the CPU’s Debug Store facility. When a branch occurs, the processor appends its source and destination addresses to this buffer until either tracing is disabled or interrupted. When the ring buffer nears interrupt threshold, the Debug Store facility raises an interrupt, giving the OS an opportunity to process the collected branch records to prevent data loss. BTS therefore provides complete, long‐duration coverage of control-flow transitions, at the cost of extra memory bandwidth and the need to allocate and periodically drain the trace buffer. Although the ring buffer’s size can be customized, its memory allocation must follow specific alignment and placement requirements. Users must configure properly and retrieve its contents—either on the real time or after execution—to reconstruct a full sequence of branch records for offline analysis.

\shortsection{Intel PT}\label{background:pt}
Intel PT is a third hardware tracing mechanism that delivers instruction-level control-flow information in a highly compact, packetized format. As the CPU runs, it emits packets that encode branch-taken targets (TIP packets) and conditional branch outcomes (TNT packets), along with timing and context-switch markers. These packets are written into a ring buffer in memory, which tools like Linux perf~\cite{noauthor_perf_nodate} can capture with minimal overhead.

However, turning the raw PT packet stream into a usable control-flow trace requires substantial post-processing. A decoder must decompress each packet, translate packet fields back into concrete instruction addresses, and correlate those addresses with the exact binary image, symbol table, and load offsets in use during tracing. In effect, it must replay the recorded execution against the original executable to reconstruct basic blocks and edges. This decoding step consumes significant CPU cycles and I/O bandwidth, often exceeding the cost of the initial trace collection. As a result, full PT trace decoding can become a performance bottleneck and a memory-pressure concern, limiting its practicality for real-time monitoring or large-scale batch analysis.

\section{System}\label{system}
In this section, we present a comprehensive description of our approach to leveraging hardware tracing capabilities, addressing existing challenges and outlining our threat model and assumptions. We further describe the detailed system design of \libiht, including the specific roles and interactions between kernel-space and user-space components, their communication mechanisms, and the workflows that enable efficient control flow analysis.

\subsection{Problem Context}\label{system:problem}
In dynamic analysis~\cite{ball_concept_1999, nethercote_dynamic_2004}, a fundamental task is the precise inference of a program’s control flow. Accurately reconstructing CFG is critical for a broad spectrum of applications including malware detection, reverse engineering, program comprehension, vulnerability assessment, and debugging. Control flow information elucidates both the structural and behavioral properties of a program, enabling analysts to identify executed paths, isolate anomalous or malicious behaviors, and decode intricate logic within binaries lacking source code. This capability is especially crucial in security contexts, where the reconstruction of CFG can expose obfuscated code, hidden functionalities, and subtle malicious actions that may evade conventional static or cursory dynamic analysis.

As discussed in \autoref{background:software-analysis}, existing methodologies exhibit significant limitations, underscoring the need for alternative approaches that deliver high-fidelity control flow reconstruction while concurrently minimizing performance overhead and resisting evasion.

\subsection{Threat Model}\label{system:threat-model}
Our threat model considers adversaries who have complete control over the distributed binaries and actively deploy sophisticated anti-analysis techniques to protect the internal logic of their applications from external scrutiny. In practice, adversaries typically attempt to detect and resist traditional software instrumentation through various defensive measures, including debugger detection, extensive code obfuscation, runtime integrity checks, and timing or environmental based evaluations specifically designed to identify performance overhead associated with instrumentation~\cite{polychronakis_measuring_2017, filho_evasion_2022, delia_evaluating_2022, delia_sok_2019, zhechev_security_2018, park_detection_2024, filho_reducing_2020}. Additionally, adversaries may employ dynamic modifications of the program's control flow at runtime, further complicating analysis efforts by obfuscating execution paths and obscuring true program behavior.

In defining our threat model, we specifically assume that adversaries neither intend nor possess the ability to gain kernel-level privileges (ring-0), and we further posit that kernel integrity remains robustly safeguarded by modern OS protections. We also explicitly state that adversaries do not have the capability to successfully compromise or subvert these kernel-level safeguards. Given these clearly defined constraints, adversaries are effectively limited to user-space manipulation only. Recognizing this critical security boundary, \libiht strategically positions its essential tracing and analysis components within the kernel space, effectively leveraging the user-kernel boundary. By operating in this secure and privileged environment, \libiht significantly enhances its capability to resist aggressive adversarial attempts at evasion, distortion, or disruption of trace data integrity, thereby preserving the reliability and accuracy of analysis results.

\begin{figure*}[ht!]
    \centering
    \includegraphics[width=\textwidth]{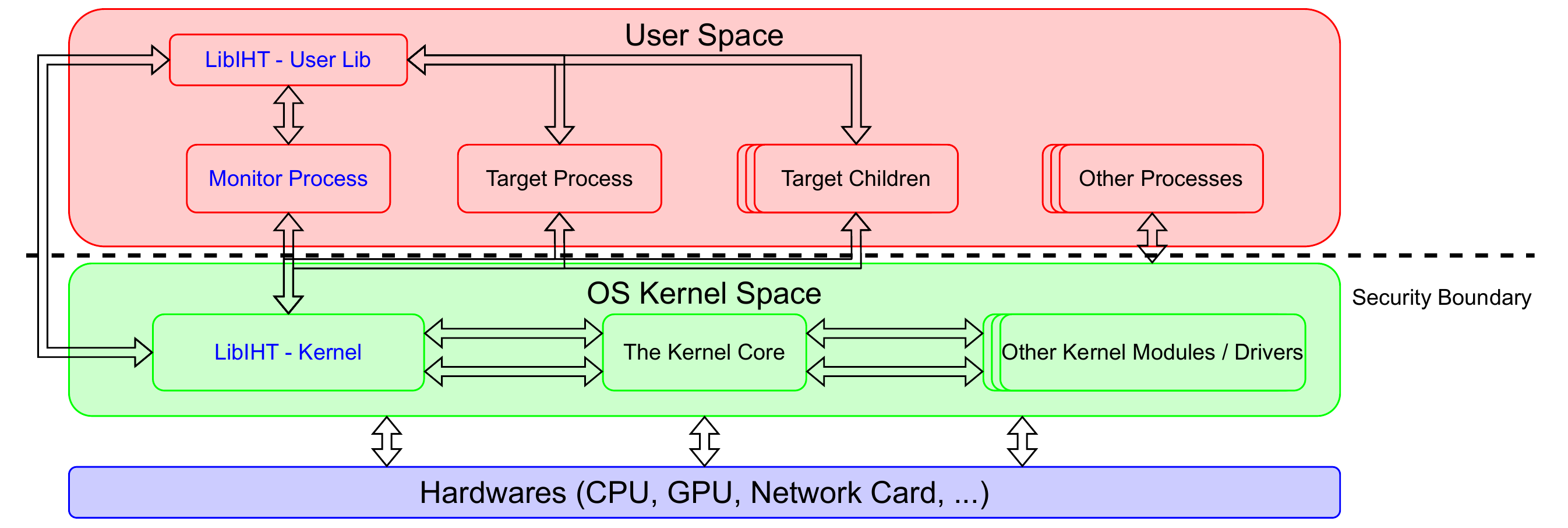}
    \caption{High-level architecture of LibIHT. The kernel-space component manages hardware tracing and collects data for the target process and its children. An optional user-space monitor (e.g., a debugger or custom tool) can analyze the trace data, though it is not required for core operation.}
    \Description{Diagram of LibIHT architecture. In user space, a monitor process and a user library interact with the target process, its children, and other processes. In kernel space, a LibIHT kernel module interfaces with the kernel core and other kernel drivers. Both user space and kernel space communicate across a security boundary, while the kernel interacts directly with hardware such as CPU, GPU, and network devices.}
    \label{fig:libiht_arch}
\end{figure*}

\subsection{System Overview}\label{system:overview}
\libiht is a dynamic binary analysis library framework that reconstructs execution control flow using hardware tracing capabilities provided by modern commodity Intel processors. It employs a dual-component architecture comprising kernel-space and user-space elements. The kernel-space component is responsible for configuring and managing hardware tracing features, while the user-space component interacts with the kernel module to retrieve and analyze trace data. \autoref{fig:libiht_arch} illustrates the high-level architecture of \libiht and interactions between these components.

\shortsection{Kernel-Space Component}\label{system:kernel}
The kernel components, implemented as Linux kernel modules and Windows kernel drivers, operate within privileged ring-0 kernel space. These modules interact directly with Intel processor hardware to access raw hardware generated trace data (LBR/BTS), manage hardware trace parameters mentioned in \autoref{background:hardware-tracing}, and securely store the generated trace data. By leveraging kernel-level privileges, \libiht ensures integrity and effectively mitigates detection and evasion attempts from user-space adversaries as studied in \autoref{system:threat-model}.

\shortsection{User-Space Component}\label{system:user}
The user components, implemented as a shared library on Linux and a dynamic link library (DLL) on Windows, offering accessible APIs for user-space applications performing security analysis. These components communicate with kernel modules through secure kernel-user interfaces, primarily using device I/O Control (\texttt{ioctl}) system calls, to retrieve raw hardware trace data. They abstract the complexity of hardware interactions by exposing high-level APIs, which process raw trace data to reconstruct control flow graphs (CFGs), thus enabling detailed security assessments. This separation between kernel and user spaces maintains strong security boundaries while optimizing usability.

\subsection{Interaction and Data Flow}\label{system:data-flow}
The operational workflow of \libiht consists of two phases: (1) Data Collection and (2) Control-Flow Reconstruction. In the Data Collection phase, a user-space component initiates a trace session by issuing \texttt{ioctl} requests to the kernel module. The kernel module then configures and activates the hardware tracing mechanisms (LBR and BTS). Once tracing is enabled, the processor begins generating branch records in real time as the program executes. The kernel module buffers these trace records in protected kernel memory, preventing any tampering by user-space malware.

In the Control-Flow Reconstruction phase, the user-space analysis library retrieves the buffered trace data via secure \texttt{ioctl} calls. The library then parses the sequence of branch records and rebuilds the program’s runtime CFG. This reconstruction involves ordering the basic blocks as they were executed and linking them into a CFG that reflects the program’s actual paths. The resulting CFG provides critical insight for analysis, enabling detection of hidden behaviors, reverse engineering of program logic, and identification of anomalies. By dividing work between kernel and user space in this manner, \libiht achieves high efficiency and remains resilient to evasion: the kernel performs low-level data capture with minimal overhead, and user-space performs heavy analysis without risking the integrity of the tracing process.

\subsection{Trade-offs and Advantages}\label{system:trade-offs}
In designing \libiht, we deliberately traded off rich semantic detail for dramatically improved performance and enhanced resistance to anti-analysis measures. Traditional software-based instrumentation tools capture comprehensive execution context-including complete instruction states, register contents, and memory access patterns-that enable highly detailed control-flow reconstruction. In contrast, \libiht relies exclusively on hardware-level branch tracing, recording only the source and destination addresses of branch events. This means that while \libiht does not capture the finer details necessary for complete state recovery, it avoids the heavy computational and storage burdens imposed by software instrumentation. For many applications, such as rapid initial reverse engineering or automated malware triage, the overall control-flow information is sufficient to pinpoint suspicious functions or anomalous behavior.

We conceptualize this design decision within a trade-off space defined by two axes: \textit{performance} and \textit{semantic granularity}. On the performance axis, software-based tools often slow down execution by orders of magnitude-our preliminary experiment results show slowdowns exceeding 100× in some cases-rendering them impractical for real-time or large-scale analysis. By shifting instrumentation to the hardware level, \libiht achieves significantly higher throughput, facilitating rapid analysis and minimally perturbing the program's natural execution flow. On the semantic granularity axis, the cost of this performance gain is the loss of detailed context: \libiht captures only branch addresses, omitting detailed state information that might be essential for deep forensic analysis. However, for early-stage reverse engineering and dynamic analysis tasks, this level of detail is typically sufficient to flag potential security issues. Importantly, these approaches are complementary. Once \libiht identifies areas of interest in an execution trace, researchers can then employ detailed software-based instrumentation or static analysis techniques to perform deeper investigation.

Moreover, by operating within kernel space and harnessing intrinsic processor features, \libiht is inherently more robust against anti-instrumentation evasion techniques. In summary, \libiht represents a balanced solution that prioritizes rapid, low-overhead trace collection and stealth while accepting a reduction in semantic detail-a trade-off that we argue is both reasonable and practical for a broad range of security and debugging applications.

\section{Evaluation}\label{eval}

We apply our approach to answer the following research questions:
\begin{itemize}
    \item \textbf{RQ1 (Benign Setting - Accuracy):} How closely does \libiht's traced CFG match actual execution paths, and what semantic details are lost when relying on hardware-based tracing instead of software instrumentation? (\autoref{eval:accuracy}) \label{rq:accuracy}
    \item \textbf{RQ2 (Benign Setting - Performance)} How does the performance of \libiht compare to traditional dynamic analysis frameworks (e.g., Intel Pin and DynamoRIO) in benign scenarios across diverse binaries? (\autoref{eval:performance}) \label{rq:performance}
    \item \textbf{RQ3 (Adversarial Setting - Resilience):} Under adversarial conditions (e.g., anti-analysis or adaptive attacks), how effectively does \libiht preserve the accuracy and performance established in RQ1 and RQ2? What techniques can adversaries use to target hardware-based tracing, and how does \libiht mitigate such attacks? (\autoref{eval:resilience}) \label{rq:resilience}
\end{itemize}

\subsection{Experimental Setup}\label{eval:setup}
We conducted our experiments on an Intel NUC 11 Essential Kit (model NUC11ATKPE), equipped with an Intel Pentium Silver N6005 processor, 8 GB of memory, and a 512 GB SSD. Ubuntu 24.04 LTS was installed to provide a modern Linux environment with up-to-date kernel support. This system was chosen because it is a commodity Intel platform that offers LBR and BTS capabilities, making it well-suited for evaluating hardware-based control-flow tracing under real-world conditions.

\shortsection{Software Baselines}\label{eval:software-baselines}
In our experiments, we compare \libiht against two widely used state-of-the-art dynamic binary analysis tools as discussed in ~\autoref{background:software-analysis}: DynamoRIO (v 11.3.0)~\cite{bruening_efficient_2004} and Intel Pin (v 3.31)~\cite{luk_pin_2005}. These tools serve as a reference baseline for evaluating \libiht on both the accuracy and performance overhead.

\shortsection{Benign Benchmark}\label{eval:benign-benchmarks}
To comprehensively assess accuracy and performance overhead under normal execution conditions, we construct a benign benchmark suite combining standard, widely-recognized programs. We include several GNU Core Utilities (Coreutils) programs, such as \texttt{ls}, \texttt{grep}, and \texttt{cat}, among others. These utilities are chosen due to their widespread use in daily computing tasks, making them ideal candidates for assessing tracing accuracy and overhead under realistic, routine user-space execution scenarios. Additionally, their control-flow diversity—from simple, linear execution to complex branching and loops—facilitates rigorous assessment of trace completeness and accuracy.

The benign benchmark form a comprehensive set of program for evaluating \libiht's tracing fidelity and instrumentation overhead in controlled, benign conditions.

\begin{table*}[ht]
\centering
\caption{Overview of Adversarial Benchmark Samples Adopted from~\cite{zhechev_security_2018}}
\label{tab:adversarial-benchmarks}
\begin{tabular}{l l l p{0.45\textwidth}}
\toprule
\textbf{Sample} & \textbf{Category} & \textbf{Subcategory} & \textbf{Technique/Description} \\ 
\midrule
\texttt{fl\_fsbase.c}    & Indirect Evasion & FL & FSBase Integrity Detection: detects discrepancies (e.g., comparing \texttt{rdfsbase} vs.\ \texttt{prctl}). \\
\texttt{fl\_ripsyscall.c}& Indirect Evasion & FL & RIP Preservation Detection: verifies that the instruction pointer remains unchanged after system calls. \\
\hline
\texttt{ca\_nx.c}        & Direct Evasion & CA & NX Page Execution Detection: tests for code cache artifacts via non-executable page checks. \\
\texttt{ca\_smc.c}       & Direct Evasion & CA & Self-Modifying Code Detection: checks handling of dynamic code modifications. \\
\texttt{ca\_vmleave.c}   & Direct Evasion & CA & VMLeave Pattern Detection: probes instrumentation layers for abnormal transitions. \\
\hline
\texttt{ea\_envvar.c}    & Direct Evasion & EA & Environment Variable Detection: searches for DBI-specific env vars (Pin, DynamoRIO, etc.). \\
\texttt{ea\_mapname.c}   & Direct Evasion & EA & Mapped File Name Detection: inspects \texttt{/proc/self/maps} for known instrumentation tools. \\
\texttt{ea\_pageperm.c}  & Direct Evasion & EA & Page Permission Detection: identifies suspicious read-write-execute memory regions. \\
\hline
\texttt{ro\_jitbr.c}     & Direct Evasion & RO & JIT Branch Timing Detection: measures anomalies in repeated code paths to detect JIT overhead. \\
\texttt{ro\_jitlib.c}    & Direct Evasion & RO & JIT Library Loading Detection: probes behavior of dynamic linking or JIT engines for abnormal latency. \\
\bottomrule
\end{tabular}
\end{table*}

\shortsection{Adversarial Benchmark}\label{eval:adversarial-benchmarks}
To evaluate the resilience and robustness of \libiht in adversarial environments, we construct a carefully curated adversarial testbench consisting of specifically chosen malware Proof-of-Concept (PoC) samples, each featuring known anti-analysis measures~\cite{zhechev_security_2018}. The adversarial benchmarks are selected to represent a broad spectrum of anti-instrumentation and evasive techniques encountered in real-world malicious software~\cite{filho_evasion_2022, delia_evaluating_2022, polychronakis_measuring_2017} including:

\shortsection{Indirect Evasion Techniques}
Indirect evasion techniques detect the side effects of instrumentation by examining abnormal processor state or wrong simulation of real instruction execution. For example, the \textbf{Functional Limitations (FL)} tests include:
\begin{itemize}
    \item \textbf{FSBase Integrity Check:} Verifies consistency in the \texttt{fsbase} register by comparing outputs from \texttt{rdfsbase} and \texttt{prctl}.
    \item \textbf{RIP Preservation Check:} Ensures that the instruction pointer remains unchanged after operations like system calls.
\end{itemize}

\shortsection{Direct Evasion Techniques}
Direct evasion techniques actively probe for artifacts introduced by instrumentation. These methods are typically subdivided as follows:
\begin{itemize}
    \item \textbf{Code Cache Artifact Detection (CA):} Techniques such as NX Page Execution Detection, Self-Modifying Code (SMC) Detection, and VMLeave Pattern Detection are used to expose shortcomings in handling dynamic code modifications or cached instrumentation artifacts.
    \item \textbf{Environment Artifact Detection (EA):} These methods search for instrumentation-related artifacts, such as environment variables commonly set by DBI frameworks, analysis of mapped files (e.g., inspecting \texttt{/proc/self/maps} for known instrumentation libraries), and detection of abnormal memory page permissions.
    \item \textbf{Runtime Just-in-Time Compiler Overhead Detection (RO):} This subcategory involves measuring execution timing anomalies across repeated code path iterations or inspecting dynamic linking behavior to detect irregularities indicative of JIT-based instrumentation.
\end{itemize}

Collectively, these adversarial benchmarks allow us to rigorously evaluate \libiht's resilience under sophisticated evasion attempts. \autoref{tab:adversarial-benchmarks} provides an overview of the adversarial test samples currently employed in our evaluation, along with a brief description of the corresponding techniques.

\shortsection{Metrics}\label{eval:trace-analysis}
After collecting traces, we evaluate \libiht along three dimensions:

\begin{itemize}
    \item \textit{Accuracy of Tracing:} We compare the CFG reconstructed by \libiht to a ground-truth CFG obtained via static analysis for each program. We use graph similarity metrics Jaccard index and normalized graph edit distance (GED) to quantify structural similarity, and we calculate coverage metrics (the percentage of basic blocks and edges from the ground truth that dynamic analysis tools captured).
    \item \textit{Performance Overhead:} We measure runtime overhead by comparing execution time with \libiht tracing against native execution time (no tracing). We report the slowdown factor and record throughput in instructions per second (IPS) as a secondary performance metric.
    \item \textit{Resilience to Evasion:} We observe whether each adversarial sample successfully detects or evades analysis. Specifically, we note if a program crashes or alters its behavior under \libiht tracing. If \libiht is partially evaded, we measure the drop in tracing accuracy and increase in overhead relative to benign conditions.
\end{itemize}

These metrics and criteria explained in the following subsections will determine how well \libiht answers our research questions.

\subsection{Accuracy in Benign Settings}\label{eval:accuracy}
In this section, we aim to answer \textbf{RQ1}, i.e., How accurately do \libiht’s CFG traces reflect true execution paths, and what semantic details are lost by using hardware-based tracing versus software instrumentation?

To systematically evaluate the accuracy of \libiht and other dynamic analysis tools, we compare their reconstructed control‐flow graphs (CFGs) against a ground‐truth CFG obtained via static analysis (using angr~\cite{shoshitaishvili_sokstate_2016}). Because dynamic tools emit only a sequence of branch addresses (i.e., a trace of basic‐block transitions), we first project that trace back onto the full, static CFG to isolate the exact execution path. Once we have two subgraphs, the ground-truth execution path and the traced path, we quantify their similarity and coverage with the following detailed metrics:

\begin{itemize}
    \item \textit{Jaccard Index:} Measures the overlap between the two subgraphs’ node and edge sets, giving a normalized score of similarity between 0 and 1. A higher Jaccard index indicates a larger shared subgraph relative to the union of both, capturing both similarity and diversity in a single value.
    \item \textit{Normalized GED:} Counts the minimum number of structural edits (node/edge insertions or deletions) required to transform the traced subgraph into the ground‐truth subgraph, then divides by the total number of elements in the ground truth. By normalizing to graph size, we account for varying trace complexities and ensure the metric remains comparable across programs of different scales.
    \item \textit{Basic Block Coverage:} The fraction of ground-truth basic blocks that appear in the traced CFG. This directly measures how comprehensively a tool records every executed block, reflecting path completeness at the node level.
    \item \textit{Edge Coverage:} The fraction of ground-truth control-flow edges (i.e., transitions between blocks) that are present in the traced CFG. This metric gauges how faithfully the tool preserves the actual execution semantics.
\end{itemize}

Together, these metrics provide complementary perspectives on tracing fidelity: the Jaccard index and normalized GED capture structural similarity and deviation, while block and edge coverage quantify completeness of execution path recovery. This multidimensional view helps illustrate the trade-offs between semantic detail and performance inherent in hardware-based tracing versus software instrumentation.

\begin{table}[t]
\centering
\caption{Mean Control-Flow Graph Reconstruction Metrics on Benign Benchmarks}
\label{tab:accuracy-benign}
\begin{tabular}{lccc}
\toprule
\textbf{Metric}       & \textbf{\libiht}      & \textbf{Intel Pin}   & \textbf{DynamoRIO}   \\
\midrule
Jaccard Index         & $98.36\pm0.14\%$      & $100\%$              & $100\%$              \\
Normalized GED        & $0.0231$              & $\approx0$           & $\approx0$           \\
Block Coverage        & $99.92\pm0.04\%$      & $100\%$              & $100\%$              \\
Edge Coverage         & $99.48\pm0.15\%$      & $100\%$              & $100\%$              \\
\bottomrule
\end{tabular}
\end{table}

\autoref{tab:accuracy-benign} reports the mean and variance of four key CFG reconstruction metrics: Jaccard index, normalized GED, block coverage, and edge coverage. The data was aggregated over all binaries inside benign benchmarks suites. The low standard deviations in each row indicate that there are no significant outliers: each tool behaves consistently across diverse binaries. 

Looking first at the software‐instrumentation baselines, both Intel Pin and DynamoRIO achieve perfect reconstruction (100\% Jaccard, $\approx$ 0 GED, and full block and edge coverage), reflecting their ability to instrument every executed instruction and record only the target process’s control‐flow transitions. In contrast, \libiht attains a Jaccard index of 98.36\% ($\pm$ 0.14\%) and a normalized GED of 0.0231, alongside nearly complete coverage (99.92\% $\pm$ 0.04\% of blocks and 99.48\% $\pm$ 0.15\% of edges). 

The slight gap between \libiht and the software baselines does not stem from missing user-mode transitions. Instead, it is caused by a small amount of noise that LBR and BTS record indiscriminately. These extra branches appear because the CPU executes a short sequence of instructions just as it enters and exits kernel mode for system calls or interrupts. At those moments, the privilege bit has not yet been fully set or cleared. As a result, even though \libiht disables tracing in privileged mode, it still captures those boundary transitions. This behavior is unavoidable given the hardware’s branch-recording granularity, but a simple post-processing step can remove all such kernel entries. By discarding any branch whose address falls outside the target process’s user-space region, one can restore a perfect 100\% match with the ground-truth CFG. Although this cleanup is not yet automated, \libiht’s raw traces already achieve over 99\% block and edge coverage, demonstrating sufficient precision for most analysis tasks.

\begin{takeaway}{Accuracy in Benign Settings}
    \libiht delivers near‐perfect CFG reconstruction with minimal variance, validating that its hardware‐assisted approach incurs only a modest and easily mitigated cost in metric‐level similarity compared to traditional software instrumentation.
\end{takeaway}

\begin{figure*}[!pht]
    \centering
    \includegraphics[width=\textwidth]{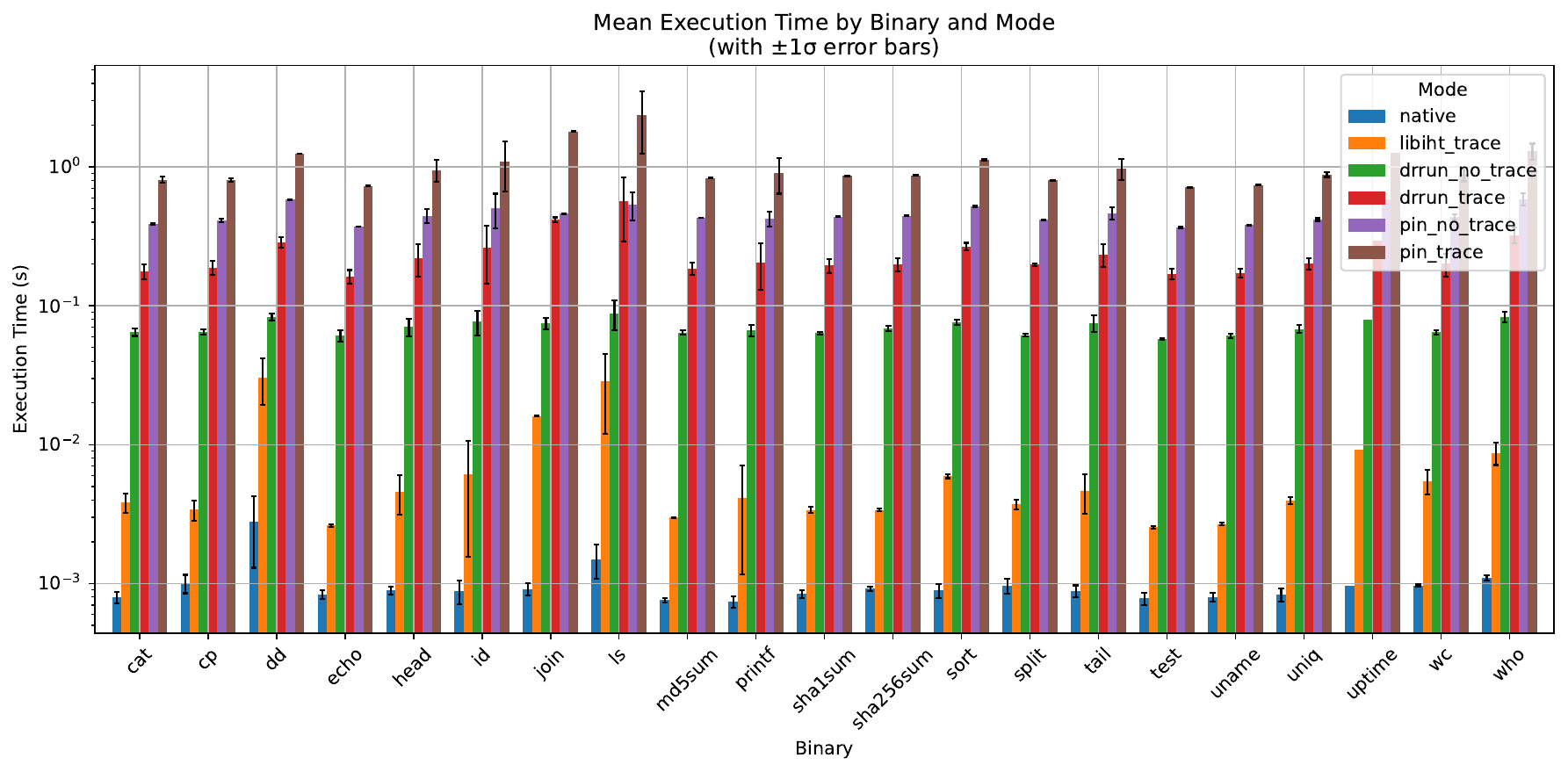}
    \caption{Comparison of mean execution times (in seconds) for standard Unix command‐line utilities under six execution modes.}
    \Description{Bar chart showing execution times of common Unix utilities across six modes: native, libiht\_trace, drrun\_no\_trace, drrun\_trace, pin\_no\_trace, and pin\_trace. Native runs are fastest, followed by libiht\_trace with small overhead. DynamoRIO and Pin add higher overhead, with traced execution being significantly slower than non-traced execution. Error bars indicate standard deviation across runs.}
    \label{fig:benign-mean-execution-time}
\end{figure*}

\begin{figure*}[!pht]
    \centering
    \includegraphics[width=\textwidth]{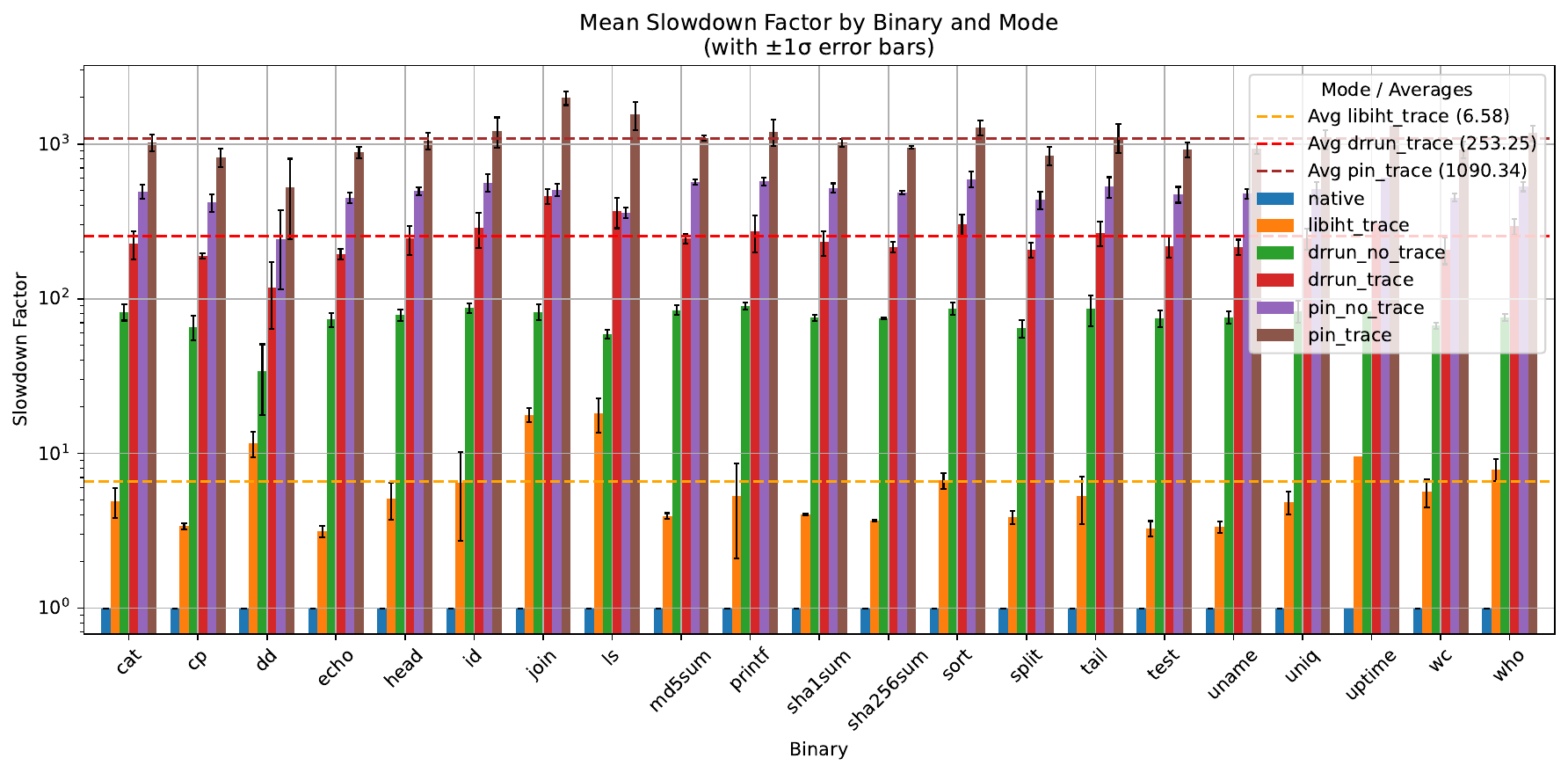}
    \caption{Average slowdown factors for standard Unix command‐line utilities relative to native execution across six modes. Dashed horizontal lines denote the overall mean slowdown across the benign benchmark}
    \Description{Bar chart of slowdown factors for Unix utilities across six modes: native, libiht\_trace, drrun\_no\_trace, drrun\_trace, pin\_no\_trace, and pin\_trace. Native execution shows baseline slowdown of 1. LibIHT trace averages around 6.6 times slower, DynamoRIO trace around 253 times slower, and Pin trace over 1000 times slower. Non-traced DynamoRIO and Pin executions still show significant overhead compared to native. Dashed lines mark mean slowdown values for each mode. Error bars represent standard deviation.}
    \label{fig:benign-mean-slowdown}
\end{figure*}

\begin{figure*}[!pht]
    \centering
    \includegraphics[width=\textwidth]{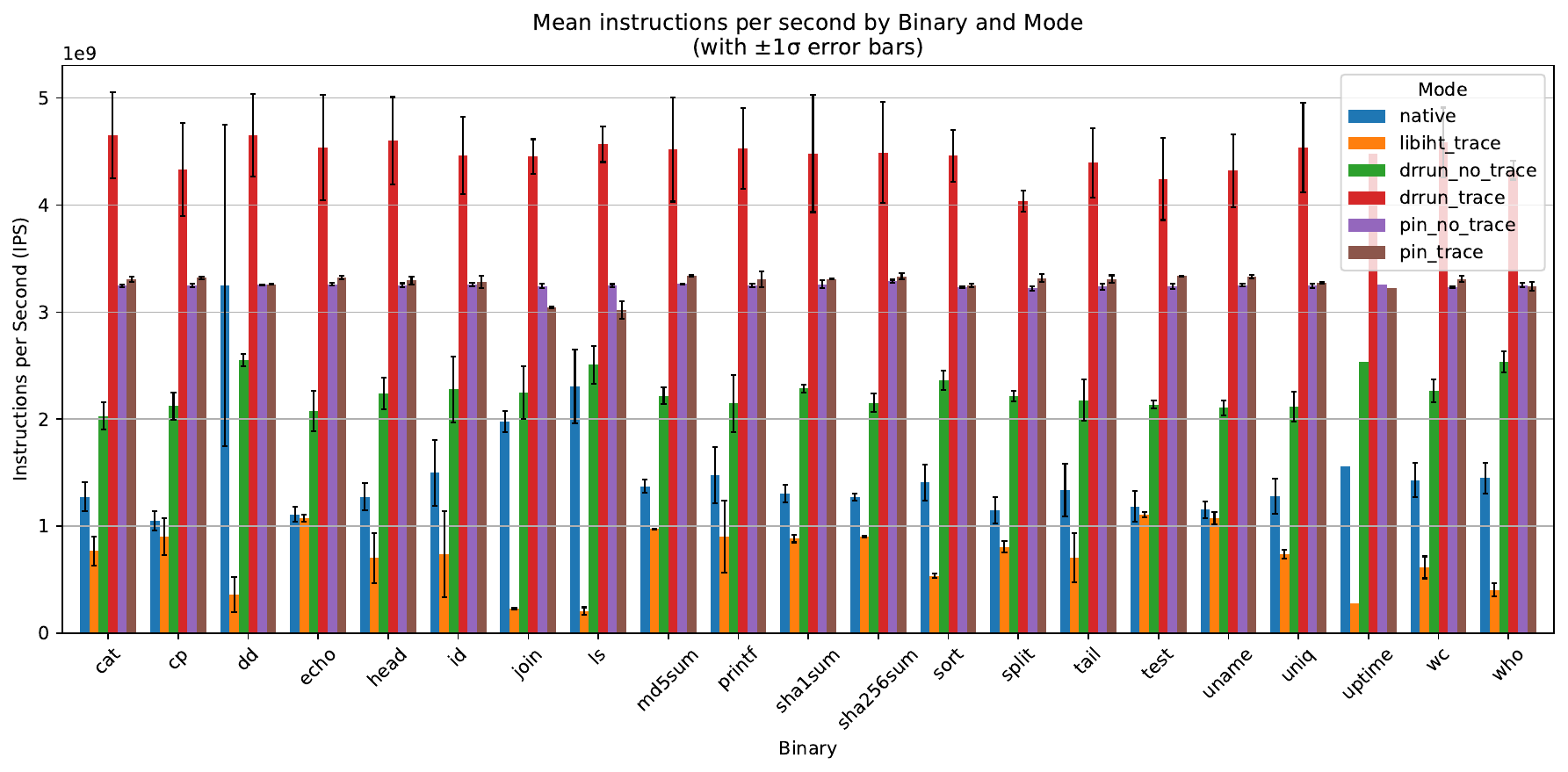}
    \caption{Average instruction per second for standard Unix command‐line utilities relative to native execution across six modes.}
    \Description{Bar chart of instructions per second across Unix utilities for six execution modes: native, libiht\_trace, drrun\_no\_trace, drrun\_trace, pin\_no\_trace, and pin\_trace. DynamoRIO with tracing shows the highest throughput, sustaining up to 4–5 billion instructions per second. Pin with and without tracing achieves slightly lower IPS around 3 billion. Native runs reach 1–3 billion depending on the utility, while libiht\_trace is lowest overall. Error bars indicate variability across runs.}
    \label{fig:benign-mean-ips}
\end{figure*}

\subsection{Performance in Benign Settings}\label{eval:performance}
In this section, we aim to answer \textbf{RQ2}, i.e., How does the performance of \libiht compare to traditional dynamic analysis frameworks (e.g., Intel Pin and DynamoRIO) in benign scenarios?

To ensure a fair and reproducible comparison, we run each binary in our benchmark suite with a small set of representative argument combinations under six modes: native execution, Intel Pin without tracing (\texttt{pin\_no\_trace}), Intel Pin with tracing(\texttt{pin\_trace}), DynamoRIO without tracing(\texttt{drrun\_no\_trace}), DynamoRIO with tracing ((\texttt{drrun\_trace})), and \libiht with tracing (\texttt{libiht\_trace}). Each configuration is executed five times per argument set, and we report the mean and standard deviation across all runs and all argument sets for each binary, thereby capturing both per‐run and per‐argument variability.\enlargethispage{12pt}

We collect two complementary performance metrics:
\begin{itemize}
  \item \textit{Execution Slowdown Factor:} Defined as 
  \[
    \mathit{Slowdown} = \frac{T_{\text{instrumented}}}{T_{\text{native}}},
  \]
  where \(T\) is the wall‐clock execution time measured by \texttt{perf}. This ratio normalizes results across different program runtimes and isolates the overhead introduced by each instrumentation framework.
  \item \textit{Instructions Per Second (IPS):} The total executed instruction count on performing the full tracing workload divided by execution time, measured by \texttt{perf stat}. IPS quantifies the impact of tracing on CPU throughput, capturing the cost of hardware logging logic (for \libiht) versus software instrumentation overhead.
\end{itemize}

By combining slowdown factors with IPS measurements, we both normalize for program length and directly observe the per‐instruction cost of each approach. This dual‐metric approach ensures that neither long‐running workloads nor instruction‐dense short workloads bias the comparison.

\autoref{fig:benign-mean-execution-time} plots the mean wall-clock time for each Unix utility on a logarithmic y-axis.  Across every benchmark, the orange \libiht bars lie very close to the blue native baseline, whereas both Intel Pin and DynamoRIO with tracing incur slowdowns of one to three orders of magnitude.  The error bars reveal that commands involving frequent user–kernel transitions—such as \texttt{printf}, \texttt{dd}, and \texttt{ls}—exhibit somewhat greater variability under \libiht.  This is because \libiht must temporarily disable and then re-enable hardware tracing around each system call or context switch, and the exact number and timing of those transitions can fluctuate with workload characteristics and OS scheduling.  In contrast, CPU-bound utilities show very tight distributions, underscoring that \libiht’s per-instruction overhead remains minimal and stable even when averaged over multiple argument runs.

Turning to \autoref{fig:benign-mean-slowdown}, we normalize each mode to native execution by plotting the slowdown factor. It reveals a consistent pattern: The orange bars (\libiht) cluster around the lower end of the scale, while the red (DynamoRIO with trace) and brown (Pin with trace) bars extend one to two orders of magnitude higher.  Dashed horizontal lines mark the mean slowdown for \libiht (6.58×), DynamoRIO (253.25×), and Pin (1090.34×), clearly illustrating that \libiht’s overhead remains roughly two orders of magnitude below that of software instrumentation for every benchmark.  This normalized view confirms the same performance advantage seen in raw execution times, without any special-casing for particular workloads.

Finally, \autoref{fig:benign-mean-ips} reports the average IPS observed while each tool executes the full tracing workload. Native runs sustain about $2.2\times10^9$ IPS, whereas \libiht records roughly $1.0\times10^9$ IPS. This gap reflects the inherent cost of on-CPU branch logging rather than any injected code. By comparison, DynamoRIO with tracing and Intel Pin without tracing report higher IPS—around $4.5\times10^9$ and $3.3\times10^9$, respectively. It is not because they execute the original program faster but due to the fact that they spend most execution cycles running instrumentation routines such as dynamic dispatch loops, just-in-time compilation, and basic-block construction. Although those routines increase the raw instruction throughput, the sheer volume of instrumentation logic also dramatically increase overall wall-clock time, as shown in \autoref{fig:benign-mean-execution-time} and \autoref{fig:benign-mean-slowdown}. In contrast, \libiht executes only the application’s native branches plus a small logging overhead, yielding much lower overall tracing cost despite a modest per-instruction slowdown.\enlargethispage{12pt}

\begin{takeaway}{Performance in Benign Settings}
    \libiht delivers dramatically lower end‐to‐end overhead compare to software‐based dynamic analysis tools. Although per‐instruction IPS is modestly reduced by hardware logging, the net performance benefit remains substantial across diverse workloads.
\end{takeaway}

\begin{table}[ht]
\centering
\caption{Detection results of dynamic analysis tools by adversarial PoC samples employing anti‐debugging and anti‐instrumentation techniques. A check mark (\cmark) indicates the tool was detected by the sample’s detection mechanism; a cross mark (\xmark) indicates the tool was not detected.}
\label{tab:detected-instr}
\begin{tabular}{l ccc}
\toprule
\textbf{Sample} & \textbf{Intel Pin} & \textbf{DynamoRIO} & \textbf{\libiht} \\
\midrule
\texttt{fl\_fsbase.c}    & \xmark     & \cmark      & \xmark      \\
\texttt{fl\_ripsyscall.c}& \xmark     & \cmark      & \xmark      \\
\hline
\texttt{ca\_nx.c}        & \xmark     & \cmark      & \xmark      \\
\texttt{ca\_smc.c}       & \cmark     & \xmark      & \xmark      \\
\texttt{ca\_vmleave.c}   & \cmark     & \xmark      & \xmark      \\
\hline
\texttt{ea\_envvar.c}    & \cmark     & \cmark      & \xmark      \\
\texttt{ea\_mapname.c}   & \cmark     & \cmark      & \xmark      \\
\texttt{ea\_pageperm.c}  & \cmark     & \cmark      & \xmark      \\
\hline
\texttt{ro\_jitbr.c}     & \cmark     & \cmark      & \xmark      \\
\texttt{ro\_jitlib.c}    & \xmark     & \xmark      & \xmark      \\
\bottomrule
\end{tabular}
\end{table}

\begin{table*}[ht]
\centering
\caption{Accuracy and Overhead Under Adversarial Benchmarks}
\label{tab:adv_accuracy_perf}
\begin{tabular}{l ccc ccc}
\toprule
& \multicolumn{3}{c}{\textbf{Block Coverage (\%)}} & \multicolumn{3}{c}{\textbf{Slowdown Factor (×)}} \\
\cmidrule(lr){2-4}\cmidrule(lr){5-7}
\textbf{Sample} & Intel Pin & DynamoRIO & LibIHT & Intel Pin & DynamoRIO & LibIHT \\
\midrule
\texttt{fl\_fsbase.c}    & 100\%     & N/A          & 99.1\%       & 674.09     & N/A          & 1.67       \\
\texttt{fl\_ripsyscall.c}& 100\%     & N/A          & 99.3\%       & 518.58     & N/A          & 1.26       \\
\hline
\texttt{ca\_nx.c}        & 100\%     & N/A          & 98.7\%       & 677.57     & N/A          & 1.86       \\
\texttt{ca\_smc.c}       & N/A     & 100\%          & 98.8\%       & N/A     & 170.13          & 1.84       \\
\texttt{ca\_vmleave.c}   & N/A     & 100\%          & 99.9\%       & N/A     & 1352.34          & 12.76       \\
\hline
\texttt{ea\_envvar.c}    & N/A     & N/A          & 99.9\%       & N/A     & N/A          & 3.28       \\
\texttt{ea\_mapname.c}   & N/A     & N/A          & 99.9\%       & N/A     & N/A          & 5.32       \\
\texttt{ea\_pageperm.c}  & N/A     & N/A          & 99.9\%       & N/A     & N/A          & 5.41       \\
\hline
\texttt{ro\_jitbr.c}     & N/A     & N/A          & 99.7\%       & N/A     & N/A          & 1.51       \\
\texttt{ro\_jitlib.c}    & 100\%     & 100\%          & 99.3\%       & 678.28     & 158.03          & 14.52       \\
\bottomrule
\end{tabular}\vspace*{-6pt}
\end{table*}

\subsection{Resilience in Adversarial Settings}\label{eval:resilience}
In this section, we aim to answer \textbf{RQ3}, i.e., Under adversarial conditions (e.g., anti-analysis or adaptive attacks), how effectively does \libiht preserve the accuracy and performance established in RQ1 and RQ2? What techniques can adversaries use to target hardware-based tracing, and how does \libiht mitigate such attacks?

We rigorously evaluate resilience through the execution of curated malware PoC samples~\cite{zhechev_security_2018, filho_evasion_2022}, each embodying specific\break anti-instrumentation and evasive strategies. First, we systematically document the success or failure of program execution by determining whether each binary detects the presence of tracing, prematurely terminates, or otherwise misbehaves during analysis. For cases of partial tracing success, we further investigate the conditioned accuracy and overhead, assessing the degree to which \libiht retains control flow tracing accuracy compared to benign scenarios (as described in RQ1). Additionally, we quantify the performance overhead of dynamic binary analysis tools introduced under various adversarial evasion techniques, comparing these results to the overhead metrics established in benign scenarios (as described in RQ2). This comprehensive evaluation illuminates the robustness and practical reliability of \libiht, highlighting its capability to mitigate adversarial threats effectively in security-critical applications.

\autoref{tab:detected-instr} shows that \libiht’s hardware‐level tracing remains undetected across every anti‐instrumentation test, whereas Intel Pin and DynamoRIO are each caught by distinct evasion strategies: the functional‐limit checks in \texttt{fl\_fsbase.c} and \texttt{fl\_ripsyscall.c} flag only DynamoRIO, illustrating that probes targeting altered processor state affect solely software DBI frameworks; among the code‐cache artifact tests, \texttt{ca\_nx.c} (NX‐page check) again detects DynamoRIO, while \texttt{ca\_smc.c} and \texttt{ca\_vmleave.c} (self‐modifying code and VM‐exit patterns) detect only Pin, reflecting tests that exploit Pin’s instrumentation model; all three environment artifact checks (\texttt{ea\_envvar.c}, \texttt{ea\_mapname.c}, and \texttt{ea\_pageperm.c}) successfully detect both Pin and DynamoRIO—by identifying DBI‐specific environment variables, mapped library names, or abnormal page permissions—but never \libiht; and in the runtime overhead probes, \texttt{ro\_jitbr.c} (JIT branch‐timing) triggers both Pin and DynamoRIO whereas \texttt{ro\_jitlib.c} (JIT‐library loading) fails to detect any tool, suggesting either unexercised detection logic or insufficient coverage.  No adversarial sample ever flags \libiht, confirming that existing anti‐analysis techniques—designed to recognize software‐injected hooks, code‐cache artifacts, or runtime overhead—are ineffective against hardware‐native tracing.

\autoref{tab:adv_accuracy_perf} then reports, for the samples where each tool successfully executed, two key metrics: block coverage and slowdown factor.  The left half of the table lists the percentage of ground‐truth basic blocks captured by each tool—Intel Pin and DynamoRIO maintain nearly 100\% coverage when they run, and \libiht likewise recovers over 98\% of blocks across all tests.  The right half shows the \nobreak slowdown factor: Pin slows execution by hundreds to over a thousand times, DynamoRIO by tens to hundreds of times, whereas \libiht’s slowdown remains below 15× in every scenario.  Even in the worst case (\texttt{ro\_jitlib.c}, where context‐switch noise increases overhead to 14.5×), \libiht outperforms the software baselines by more than an order of magnitude.  

\begin{takeaway}{Resilience in Adversarial Settings}
    \libiht not only evades all common anti‐instrumentation checks but also retains high CFG coverage and orders‐of‐magnitude better performance under adversarial conditions—demonstrating its robustness and efficiency as a hardware‐assisted tracing solution.
\end{takeaway}

\section{Discussion}\label{discussion}
The evaluation results suggest that hardware-assisted tracing via \libiht is a viable and even preferable alternative to traditional software instrumentation in many scenarios. In this section, we reflect on the implications of these findings, discuss the limitations of our approach, and outline opportunities for future work.

\shortsection{Benefits and Implications}
An immediate takeaway is that the reduction in overhead (an order of magnitude less than Pin or DynamoRIO) can unlock new uses for dynamic analysis. Techniques that were previously impractical on resource-constrained or real-time systems might become feasible with \libiht. Moreover, the resilience demonstrated against anti-analysis measures implies that \libiht could be particularly useful in malware analysis pipelines, where stealth is paramount. By shifting the tracing to hardware and kernel space, we significantly lower the “interference profile” of the analysis tool, which is a new point in the design space for program analysis tools.\enlargethispage{12pt}

\shortsection{Limitations}
Despite these advantages, \libiht is not without trade-offs. One limitation is the loss of certain semantic information - for instance, \libiht currently logs control-flow branches but not the associated data values or memory accesses. This means analyses that require fine-grained data flow or taint tracking cannot be directly supported in our framework. Another limitation is platform dependence: our implementation relies on Intel-specific features (LBR/BTS). While many modern CPUs have analogous capabilities, additional engineering is required to support other architectures (e.g., ARM’s branch record features). \libiht can, in principle, capture control flows even within self-modifying or dynamically generated code. However, because such code is not statically available, the raw traces alone may have limited utility for higher-level analyses. In practice, meaningful interpretation of these traces may still require complementary techniques such as selective software instrumentation. A further consideration is that, although \libiht’s use of LBR/BTS resists all known software anti-instrumentation checks, attackers could eventually target hardware tracing directly-e.g., by overflowing the LBR buffer, exploiting microarchitectural quirks to inject spurious branches, or detecting MSR-access timing anomalies. Over time, such evasion techniques could erode \libiht’s stealth advantage unless countermeasures (e.g., randomized drain intervals or noise injection) are adopted. We leave the exploration of such targeted attacks to future work.\enlargethispage{12pt}

\shortsection{Future Work}
There are several avenues to extend this work. First, enriching trace semantics is a priority: we plan to investigate combining \libiht with lightweight instrumentation that logs selective additional context (such as function call arguments or memory addresses) to recover some of the lost semantic detail without reintroducing high overhead. Second, porting \libiht to other platforms (ARM, AMD processors) and evaluating it there would broaden its applicability. Third, a deeper integration with analysis tools (like feeding \libiht’s output into existing CFG recovery or taint analysis frameworks) could demonstrate end-to-end use cases. Finally, user studies or case studies (e.g., analyzing real malware samples in the wild) would help validate \libiht’s effectiveness in practical security workflows.

Another promising direction is the use of \libiht for fuzzing. Modern hardware-assisted fuzzers have employed Intel PT to collect execution coverage efficiently, enabling large-scale exploration of complex programs~\cite{schumilo_kaflhardware-assisted_2017, zhang_ptfuzz_2018, chen_ptrix_2019}. However, PT’s trace decoding overhead and complexity can introduce bottlenecks in coverage-guided fuzzing workflows. \libiht, by offering simpler and lower-overhead control-flow tracing, may serve as an attractive alternative. Future work could benchmark \libiht against Intel PT-based fuzzers to evaluate trade-offs in accuracy, throughput, and scalability, and to assess whether \libiht can expand the practicality of hardware-assisted fuzzing in security-critical domains.

Overall, our discussion highlights that \libiht’s hardware-centric approach shifts some long-standing trade-offs in dynamic analysis. It achieves a new balance between performance, transparency, and fidelity. While there are limitations to address, the approach opens up a promising direction for building low-overhead, anti-evasion program analysis tools.
\section{Related Work}\label{related-work}
In this section, we provide an in-depth review of prior research that utilizes hardware tracing features for program analysis. We categorize the literature into two primary streams: (i) Intel PT-based approaches and (ii) LBR/BTS-based approaches. The following discussion details the advantages and limitations of these approaches and explains how they motivate the design of \libiht.\enlargethispage{12pt}

\subsection{Intel PT-Based Approach}\label{related-work:pt}
Intel has emerged as a powerful mechanism for capturing fine-grained execution information. Several tools have integrated Intel PT into their frameworks, such as Linux Perf~\cite{noauthor_perf_nodate} for performance analysis, and specialized systems for control flow integrity monitoring~\cite{gu_pt-cfi_2017, ge_griffin_2017}. Intel PT offers high-precision, instruction-level tracing that enables detailed reconstruction of program execution paths-a valuable feature for debugging and reverse engineering.

Despite these strengths, Intel PT-based systems face significant challenges. The voluminous trace data generated requires extensive storage and imposes heavy computational overhead during decoding and analysis. This high data granularity often leads to performance bottlenecks, making real-time analysis difficult, especially on resource-constrained platforms. Although researchers have proposed techniques such as selective tracing and data compression~\cite{gu_pt-cfi_2017, ge_griffin_2017} to alleviate these issues, such methods frequently introduce trade-offs that compromise trace fidelity or add complexity to the analysis pipeline. Consequently, while Intel PT delivers unmatched detail, its overhead and scalability concerns restrict its practical applicability in scenarios requiring low-latency or continuous monitoring.

\subsection{LBR and BTS-Based Research}\label{related-work:lbr-and-bts}
An alternative line of research has focused on leveraging the more lightweight hardware features available in modern processors, the LBR and BTS. LBR provides a circular buffer that records the most recent branch transitions, offering a succinct snapshot of control-flow events. BTS extends this capability by logging branch events into a dedicated memory buffer over a longer period. These mechanisms naturally produce substantially less data compared to Intel PT, thereby reducing storage and processing overhead.

Prior studies have applied LBR/BTS-based tracing across various domains. For example, Willems et al.~\cite{willems_down_2012} employed LBR for malware analysis by detecting anomalies in branch patterns, while other works have explored its use for exploit mitigation~\cite{cheng_ropecker_2014, pappas_transparent_2013, zhou_hardstack_2019, liu_retrofitting_2022} and performance profiling~\cite{arulraj_leveraging_2014, marin_break_2021}. Although these efforts demonstrate that processor-based tracing can effectively capture control-flow anomalies, many existing solutions remain narrowly focused, addressing only specific security threats or performance issues in isolated settings. Moreover, several approaches are presented merely as proof-of-concept prototypes, lacking the robustness and scalability needed for general-purpose program analysis. Common limitations include fixed buffer sizes, constrained tracing durations, and challenges in integrating with binary analysis tools.

\libiht addresses these limitations by providing a unified, modular framework that integrates both LBR and BTS tracing. Unlike prior approaches that are either tailored to specific applications or exist solely as isolated prototypes, \libiht offers a robust kernel-space implementation paired with flexible user-space APIs. This design enables efficient branch-level trace capture with low overhead while facilitating seamless integration with diverse security and debugging workflows. In doing so, \libiht overcomes the issues of scalability, inflexibility, and high resource consumption that have limited earlier tools, thereby advancing the state of the art in hardware-assisted dynamic analysis.
\section{Conclusion}\label{conclusion}
In this paper, we presented \libiht, a hardware-assisted dynamic analysis framework that leverages CPU branch tracing features to achieve efficient and stealthy program tracing. \libiht directly addresses the limitations of software-based instrumentation: in our experiments, it delivered order-of-magnitude performance improvements and resisted a range of anti-analysis techniques, all while accurately reconstructing program control-flow. By relocating tracing into the processor and kernel, \libiht minimizes its footprint in the target’s execution, thereby avoiding the perturbations and detections that plague traditional tools. \enlargethispage{12pt}

Our work contributes a novel perspective on program analysis—one that exploits modern hardware capabilities to balance the long-standing unexplored trade-off between analysis fidelity and overhead. The positive results from \libiht’s evaluation suggest that hardware-centric tracing can be a practical foundation for building next-generation analysis tools. We believe this approach will enable security researchers and software engineers to analyze programs (including malware) in scenarios that were previously too slow or too risky using conventional methods. As hardware support for tracing continues to evolve, frameworks like \libiht can be extended and adapted to further close the gap between transparent, high-fidelity analysis and real-world performance constraints.

\begin{acks}
\textbf{Funding acknowledgment:}
This material is based upon work supported by the National Science Foundation under Grant No. CNS-2343611. Any opinions, findings, and conclusions or
recommendations expressed in this material are those of the author(s) and do not
necessarily reflect the views of the National Science Foundation.
This work was supported in part by the Semiconductor Research Corporation (SRC) and DARPA.

\end{acks}

\bibliographystyle{ACM-Reference-Format}
\bibliography{refs}

\appendix{}
\section{Appendix}\label{appendix}

\subsection{Workload Commands for Motivating Experiment}\label{appendix:workload-commands}
In \autoref{tab:instr_overhead}, we simply measure the runtime overhead introduced by dynamic binary instrumentation tools, we selected a set of commonly used \texttt{coreutils} programs that represent typical system workloads, including I/O, file system metadata access, and CPU-bound operations. Each command was chosen to be deterministic and environment-stable.

The exact commands used in the evaluation are listed below:

\begin{itemize}
    \item \texttt{ls}: \texttt{ls -lah /bin > /dev/null}
    \item \texttt{dd}: \texttt{dd if=/dev/zero of=/dev/null bs=1M count=100}
    \item \texttt{echo}: \texttt{echo 'hello world'}
    \item \texttt{sort}: \texttt{sort < /etc/passwd > /dev/null}
    \item \texttt{wc}: \texttt{wc -l < /etc/passwd > /dev/null}
    \item \texttt{cat}: \texttt{cat /etc/passwd > /dev/null}
\end{itemize}

All output from these commands was suppressed to ensure that measurement reflected only core execution time, without terminal I/O interference. Each workload was run in three configurations: native, under Intel Pin (\texttt{pin -- <command>}), and under DynamoRIO (\texttt{drrun -- <command>}).

\end{document}
\endinput{}